# Capture and light-induced release of antibiotics by an azo dye polymer.

Stephen Atkins[1], Alysa Chueh[2], Taylor Barwell[1], Jean-Michel Nunzi[2,3] and Laurent Seroude[1]

*Department of Biology[1], Department of Chemistry[2], Department of Physics, Engineering Physics and Astronomy[3], Queen's University, Kingston, ON, Canada*

## Abstract

The isomerisation of azo dyes can induce conformational change which have potential applications in medicine and environmental protection. We developed an agar diffusion assay to test the capture and release of biologically active molecules from an azo electro-optic polymer, Poly (Disperse Red 1 methacrylate) (DR1/PMMA). The assay monitors the growth of bacteria placed in soft agar under a glass coverslip. Antibiotics can then be applied on the coverslip resulting in the clearance of the area under the coverslip due to growth inhibition. This assay demonstrates that DR1/PMMA is able to capture either tetracycline or ampicillin and the relative amount of DR1/PMMA required for capture was determined. Finally, the active antibiotics can be released from DR1/PMMA by exposure to green light but not by exposure to white light or heat.

## Introduction

Materials containing azo-benzene chromophores exhibit a range of photo-responsive properties[1]. Under various light irradiation, -N=N- azo groups change from their *trans* to their *cis* conformations. The area of motions generated by azobenzene isomerization in polymer materials range from slight reorientations of the chromophore to mass motion of the polymer as well as, surface relief grating, and a variety of nonlinear optical effects[2]. Poly (Disperse Red 1 methacrylate) (DR1/PMMA) is a polymer with a -N=N- azo group attached to the backbone MMA monomers through covalent bonds. DR1/PMMA can occur in two different geometries, a linear stretched *trans* configuration and an angular *cis* conformation. The polymer exhibits good thermal and temporal stabilities with a high glass transition temperature[3], $T_g = 91\ ^oC$[4].

In this report, a biological assay is designed to test the ability of DR1/PMMA to capture and release biologically active molecules. Two broadly used antibiotics, ampicillin and tetracycline are tested. Tetracycline is the simplest member of the "tetracycline" class of antibiotics defined by the presence of the DCBA naphthacene core containing four aromatic rings[5]. Tetracycline binds to the 30S ribosomal subunit and leads to the inhibition of protein translation by preventing tRNAs docking. Ampicillin belongs to the beta-lactam class of antibiotics defined by the presence of a highly reactive 3-carbon and 1-nitrogen ring[6]. Ampicillin interferes with the synthesis of peptidoglycan necessary for the bacterial cell wall. Ampicillin and tetracycline are polar molecules that can form hydrogen bonds with DR1/PPMA.

## Results

**Agar diffusion assay**

An agar diffusion assay was first developed to be able to monitor the growth of bacteria placed in soft agar under a glass coverslip. Four Escherichia coli strains commonly used in molecular biology were used (Figure S1). The presence of tetracycline and ampicillin will be respectively detected with the tetracycline-sensitive (tet$^s$) TB1 and HB101 strains and ampicillin-sensitive (amp$^s$) DH5α and XL1Blue strains. TB1 and HB101 have been transformed with the pUAST and pBSGFP plasmids conferring resistance to ampicillin. XL1Blue and DH5α respectively contains a F' factor and the pHC60 plasmid conferring resistance to tetracycline. The resistance to tetracycline and ampicillin eliminates the occurrence of antibiotic-sensitive contaminants and facilitates the assay for experimentalists unacquainted with aseptic techniques. Additionally the tet$^s$ amp$^r$ HB101:pBSGFP and the amp$^s$ tet$^r$ DH5α:pHC60 strains produce GFP for alternative visualization by fluorescence. Because the assay will be standardized by inoculating the soft agar medium with a fixed amount of a liquid culture, the growth of each strain in a liquid culture was examined by measuring the optical density at 600nm. Since the optical density is affected by cell size and cannot differentiate between dead and alive cells, the number of cells alive was measured simultaneously by serial dilution plating.

Next soft agar plates were inoculated with different amounts of liquid cultures of each strain and the bacterial growth occurring under a clean glass coverslip was qualitatively assessed for the easiest visualization (Figure 1). The plates with higher cell density clearly show that the biggest colonies grew on the surface of the top agar while the colonies below the surface were smaller and the smallest colonies were always found under the coverslip as expected from the decreased oxygen availability. As the cell density decreases, the colonies under the coverslip have obviously increased their size and are now similar to the colonies outside of the coverslip thereby greatly improving visualization with the naked eye.

Next the amount of antibiotic required to inhibit bacterial growth was determined. 10 µl aliquots of ampicillin or tetracycline dissolved in dimethylformamide (DMF) at appropriate concentrations were pipetted on clean glass coverslips and the solvent was evaporated. This process generated series of coverslips with different amounts of antibiotics that were then tested on the bacterial strains (Figure 2). The tet$^s$ strains show that 0.313 µg of tetracycline is the lower amount required to observe some bacterial clearance but 1.25 µg is needed for complete clearance and higher amounts extend the clearance zone well outside the coverslip area. It is noticeable that HB101 is slightly less sensitive than TB1. The amp$^s$ strains show that 1.25 µg of ampicillin is required to observe some clearance but 5 µg is needed for complete clearance. These results agree with the 12.5-25 µg/ml tetracycline and 50-100 µg/ml ampicillin concentrations routinely used to select antibiotic-resistant *E. coli*.

**DR1/PMMA can capture antibiotics and restore bacterial growth**

Although the chemical structures of ampicillin and tetracycline allow hydrogen binding with DR1/PMMA, it is not possible to mathematically determine the amount of DR1/PMMA needed since the number of antibiotic molecules that can bind per structural unit is unknown. Therefore, the ability to capture these antibiotics and the appropriate molecular ratio was determined experimentally with the agar diffusion assay. After verification that it does not affect bacterial growth (Figure S2), different amounts of DR1/PMMA were added to 0.625 µg tetracycline or 2.5 µg ampicillin and the resulting coverslips were tested (Figure 3). Compared to the control coverslips, the addition of DR1/PMMA reduced the clearance zone caused by either antibiotic. However, both antibiotics required 600 µg DR1/PMMA to fully restore bacterial growth. These observations show that DR1/PMMA is able to capture both antibiotics and capture is more efficient with ampicillin than tetracycline.

**Antibiotics can be released with green light**

Once it has been established that DR1/PMMA protects E. coli from either antibiotic, the agar diffusion assay can be used to test whether the antibiotic can be released and has retained its biological activity. DR1 isomerization is typically achieved by exposure to wavelengths between 488 nm and 532 nm[1,7]. It has been known for a long time that visible light above 400nm can affect respiration, ATPase activity and decreases ATP level of the ML 308 *E. coli* strain[8]. More recently it has been observed that exposure to green light can affect the viability and growth of the MG 1655 *E. coli* strain[9]. Therefore, the effect of green light on the agar diffusion assay was tested and found not to affect the growth of the four *E. coli* strains (Figure S3). Once exposed to green light, the DR1/PMMA+antibiotic coverslips restore the clearance zone demonstrating that both antibiotics can be released and are still active (Figure 4). The exposed DR1/PMMA+ampicillin coverslips are almost identical to the positive controls indicating complete release whereas the DR1/PMMA+tetracycline coverslips suggest that the release is incomplete or/and that tetracycline lost some activity. Importantly release is not observed when the coverslips are exposed to heat (Figure S4) or to white light (Figure S5). Although white light includes green wavelengths, the intensity is insufficient as preliminary observations indicated that the release of ampicillin is not detectable after one-hour exposure and required overnight exposure (Figure S6).

## Discussion

The light-induced release of antibiotics by the DR1/PMMA polymer can be attributed to different molecular processes. Water could dissolve into the water insoluble-polymer under illumination and wash the antibiotic out of the polymer, like DR1-molecules dissolved into a cellulosic polymer host under light[10]. However, the experiment did not show any swelling of the neat polymer under illumination. The photoinduced movement of the DR1-chromophores could also be suspected to expel the antibiotic from the host.

Molecular dynamics simulations indeed suggest that the repeated isomerization of DR1 molecules induces an accelerated diffusion of the host[11]. However, an experiment performed using a molecular DR1-glass showed that phase separation does not happen under illumination[12]. The spinodal decomposition of linear polymers is a well-documented entropy driven event[13]. We suggest that a photo-induced spinodal decomposition happens during illumination[14], which expels the antibiotic from the entangled DR1/PMMA polymer chains into the water of the assay.

## Conclusion

The ability to control the activity of biological molecules as well as the ability to relieve the control is critical for a multitude of applications in medical and environmental sciences. It is widely recognized that antibiotics are less effective mainly because of the evolution of resistant bacteria. The possibility of releasing an antibiotic only where needed allows increasing therapeutic doses without exposing the whole microbiota as well as minimizing risks of side effects. The trapping of antibiotics has the additional advantage to minimize the impact of excreted molecules and discarded pills.

The agar diffusion assay is a very simple and cheap method to quickly assess the ability to capture and release antibiotics. In this report, the assay was used to demonstrate that ampicillin and tetracycline can be captured by DR1/PMMA and can subsequently be released by exposure to green light. It appeared that the capture and release with this polymer is more efficient for ampicillin than tetracycline but the ease of the assay allows screening through any polymers or molecules.

## Methods

### Bacteria strains

XL1Blue (recA1 endA1 gyrA96 thi-1 hsdR17 supE44 relA1 lac [F' proAB lacI$^q$ Δ(lacZ)M15 Tn10 (Tet$^r$)]) has been obtained from Stratagene. TB1:pUAST (ara Δ(lac proAB) [Φ80dlac Δ(lacZ)M15] rpsL (Str$^r$) thi hsdR: pUAST (Amp$^r$)) and DH5α:pHC60 (recA1 endA1 gyrA96 thi-1 hsdR17 relA1 [Φ80dlac Δ(lacZ)M15] Δ(lacZYA argF)U169:pHC60 (Tet$^r$)) have been previously described[15,16]. HB101:pBSGFP: supE44 ara14 galK2 lacY1 Δ(gpt-proA)62 rpsL20 (Str$^r$) xyl-5 mtl-1 recA13 Δ(mcrC$^-$ mr$^r$) hsdS-(r$^-$ m$^-$):pBSGFP (Amp$^r$) was obtained by transformation of HB101 with the pBSGFP plasmid (a pBluescript plasmid expressing GFP). All strains were cultured on 2TY (1% yeast extract, 1.6% tryptone, 0.5% NaCl) agar (1.5%) plates or broth supplemented with 25 µg/ml tetracycline (tet$^r$ strains) or 100 µg/ml ampicillin (amp$^r$ strains).

### Solutions and coverslips

Poly (Disperse Red 1 methacrylate) (DR1/PMMA) (Sigma cat#579009), ampicillin (BioShop cat#AMP201) and tetracycline (BioShop cat#TET701) solutions and dilutions were prepared using dimethylformamide (DMF) (Fisher) as the solvent. Dilutions were prepared by 2-fold serial dilutions of 100 mg/ml ampicillin, 25 mg/ml tetracycline and 60 mg/ml DR1/PMMA stock solutions. 10µl of the appropriate solutions were pipetted onto circle glass coverslips (Fisher cat#12-545-101 or Ultident cat#170-C12MM) and air dried at room temperature in the dark for a minimum of 16h. Mixture solutions were obtained by mixing antibiotic and DR1/PMMA solutions or by dissolving the desired amount of DR1/PMMA in the antibiotic solution.

**Agar diffusion assay**

2TY soft agar (2TY supplemented with 7 g/l agar) was autoclaved and cooled down to 47˚C before adding ampicillin (100µg/ml final, amp$^r$ strains) or tetracycline (25µg/ml final, tet$^r$ strains), and the desired number of bacteria from a fresh or overnight culture. The optical density of cultures was measured at 600nm (always between 2 and 6) and a 100-fold dilution of the cultures were prepared. Based on the results presented in Figure 1, 6µl (TB1, HB101 and DH5α strains) or 25µl (XL1Blue strain) of the culture dilution was added per 15ml of soft agar. 15ml of the soft agar/antibiotic/bacteria solution was then poured in 10cm plastic petri dishes. Once the soft agar has jellified, coverslips were placed atop the surface. It is important not to move the coverslips once placed as it was observed that the antibiotic from control coverslips diffuse in the soft agar almost instantaneously (after a slight move of the coverslip, the growth inhibition zone remained centered to the original position as can be seen for the DH5α strain antibiotic controls in Figure 4). Plates were incubated overnight (12 to 18h) in a 37˚C bacterial incubator. Dark plates were placed inside a black box while exposed plates had a green laser light (532nm 50mw green laser diode module, eBay) positioned overtop the coverslips.

## Figure legends

**Figure 1: Influence of cell density on growth under coverslips.**

The optical density at 600nm of the liquid cultures is indicated for each bacterial strain (first picture of each row). The soft-agar was inoculated with aliquots of 1/100 dilutions of the cultures and the equivalent amount of undiluted cultures is indicated at the top of each column.

**Figure 2: Determination of the amount of antibiotic in the agar diffusion assay.**

The mass of antibiotic applied to each coverslip is indicated at the top of each column. The top two rows (TB1 and HB101 strains) used tetracycline coverslips while the bottom two rows (XL1Blue and DH5α) used ampicillin coverslips.

**Figure 3: Antibiotic capture with DR1/PMMA.**

All coverslips have been prepared with 0.625 µg tetracycline (top two rows) or 2.5 µg ampicillin (bottom two rows) supplemented with the mass of DR1/PMMA indicated at the top of each column. Inset pictures in the bottom row show alternative visualization using GFP fluorescence.

**Figure 4: Antibiotic release by exposure to green light.**

Negative control coverslips (Control) have neither antibiotic or DR1/PMMA. 0.625 $\mu$g tetracycline (top four rows) or 2.5 $\mu$g ampicillin (bottom four rows) was applied to the positive control coverslips (Antibiotic). 600$\mu$g DR1/PMMA was applied to no antibiotic control coverslips (DR1/PMMA). 0.625 $\mu$g tetracycline (top four rows) or 2.5 $\mu$g ampicillin (bottom four rows), and 600$\mu$g DR1/PMMA was applied to the experimental coverslips (DR1/PMMA + Antibiotic).

## References

1   Mazaheri, L., Bobbara, S. R., Lebel, O. & Nunzi, J. M. Photoinduction of spontaneous surface relief gratings on Azo DR1 glass. *Opt Lett* **41**, 2958-2961, doi: 10.1364/OL.41.002958 (2016).
2   Natansohn, A. & Rochon, P. Photoinduced motions in azo-containing polymers. *Chem Rev* **102**, 4139-4175 (2002).
3   Wu, X., Nguyen, T. T. N., Ledoux-Rak, I., Nguyen, C. T. & Lai, N. D. in *Holography - Basic Principles and Contemporary Applications* (ed Emilia Mihaylova) Ch. 7, 147-170 (IntechOpen, 2013).
4   Kirby, R., Sabat, R. G., Nunzi, J.-M. & Lebel, O. Disperse and disordered: a mexylamino triazine substituted azobenzene derivative with superior glass and surface relief grating formation. *J. Mater. Chem C* **2,** 841-847, doi: 10.1039/c3tc32034k (2014).
5   Nguyen, F. *et al.* Tetracycline antibiotics and resistance mechanisms. *Biol Chem* **395**, 559-575, doi: 10.1515/hsz-2013-0292 (2014).
6   Kong, K. F., Schneper, L. & Mathee, K. Beta-lactam antibiotics: from antibiosis to resistance and bacteriology. *APMIS* **118**, 1-36, doi: 10.1111/j.1600-0463.2009.02563.x (2010).
7   Fiorini, C. *et al.* Molecular migration mechanism for laser induced surface relief grating formation. *Synthetic Met* **115**, 121-125, doi: 10.1016/S0379-6779(00)00332-5 (2000).
8   D'Aoust, J. Y., Giroux, J., Baraan, L. R., Schneider, H. & Martin, W. G. Some effects of visible light on Escherichia coli. *J Bacteriol* **120**, 799-804 (1974).
9   Giannakis, S., Rtimi, S., Darakas, E., Escalas-Canellas, A. & Pulgarin, C. Light wavelength-dependent E. coli survival changes after simulated solar disinfection of secondary effluent. *Photochem Photobiol Sci* **14**, 2238-2250, doi: 10.1039/c5pp00110b (2015).
10  Pawlicka, A., Sabadini, R. C. & Nunzi J.-M. Reversible light-induced solubility of disperse red 1 dye in a hydroxypropyl cellulose matrix. *Cellulose* **25**, 2083–2090, doi: 10.1007/s10570-018-1672-z (2018).
11  Saiddine, M., Teboul, V. & Nunzi, J.-M. Isomerization-induced surface relief gratings formation: A comparison between the probe and the matrix dynamics. *J. Chem. Phys.* **133**, 044902, doi: 10.1063/1.3465577 (2010).

12  Laventure A. *et al.* Photoactive/passive molecular glass blends: an efficient strategy to optimize azomaterials for surface relief grating inscription. *ACS Appl. Mater. Interfaces* **9**, 798–808, doi: 10.1021/acsami.6b11849 (2017).

13  Xie, S., Natansohn, A. & Rochon, P., Compatibility Studies of Some Azo Polymer Blends. *Macromol.* **27**, 1489-1492, doi: 10.1021/ma00084a033 (1994).

14  Tran-Cong-Miyata, Q. & Nakanishi, H., Phase separation of polymer mixtures driven by photochemical reactions: current status and perspectives. *Polym. Int.* **66**, 213-222, doi: 10.1002/pi.5243 (2017).

15  Brand, A. H. & Perrimon, N. Targeted gene expression as a means of altering cell fates and generating dominant phenotypes. *Development* **118**, 401-415 (1993).

16  Elrod-Erickson, M., Mishra, S. & Schneider, D. Interactions between the cellular and humoral immune responses in Drosophila. *Curr Biol* **10**, 781-784 (2000).

## Author information

### Affiliations

*Department of Biology, BioSciences Complex, Queen's University, 116 Barrie St, Kingston ON K7L 3N6, Canada*
Stephen Atkins, Taylor Barwell and Laurent Seroude

*Department of Chemistry, Department of Physics, Engineering Physics and Astronomy, Chernoff Hall, Queen's University, 90 Bader Lane, Kingston ON K7L 3N6, Canada*
Alysa Chueh and Jean-Michel Nunzi

### Contributions

S.A., A.S. and T.B. participated in the design of this study, performed experiments and analyzed data. L.S. designed and oversaw the study, performed experiments, analyzed and interpreted data, and drafted the manuscript. J-M.N. participated in the design and oversaw the study, interpreted data, and drafted the manuscript. All authors read and approved the final manuscript.

### Competing interests

The authors declare no competing financial interests.

### Corresponding authors

Correspondence to Jean-Michel Nunzi (nunzijm@queensu.ca) or Laurent Seroude (seroudel@queensu.ca).

### Funding

J-M. N. acknowledges the NSERC-Canada Research Chair Program entitled *Photonics for Life.*

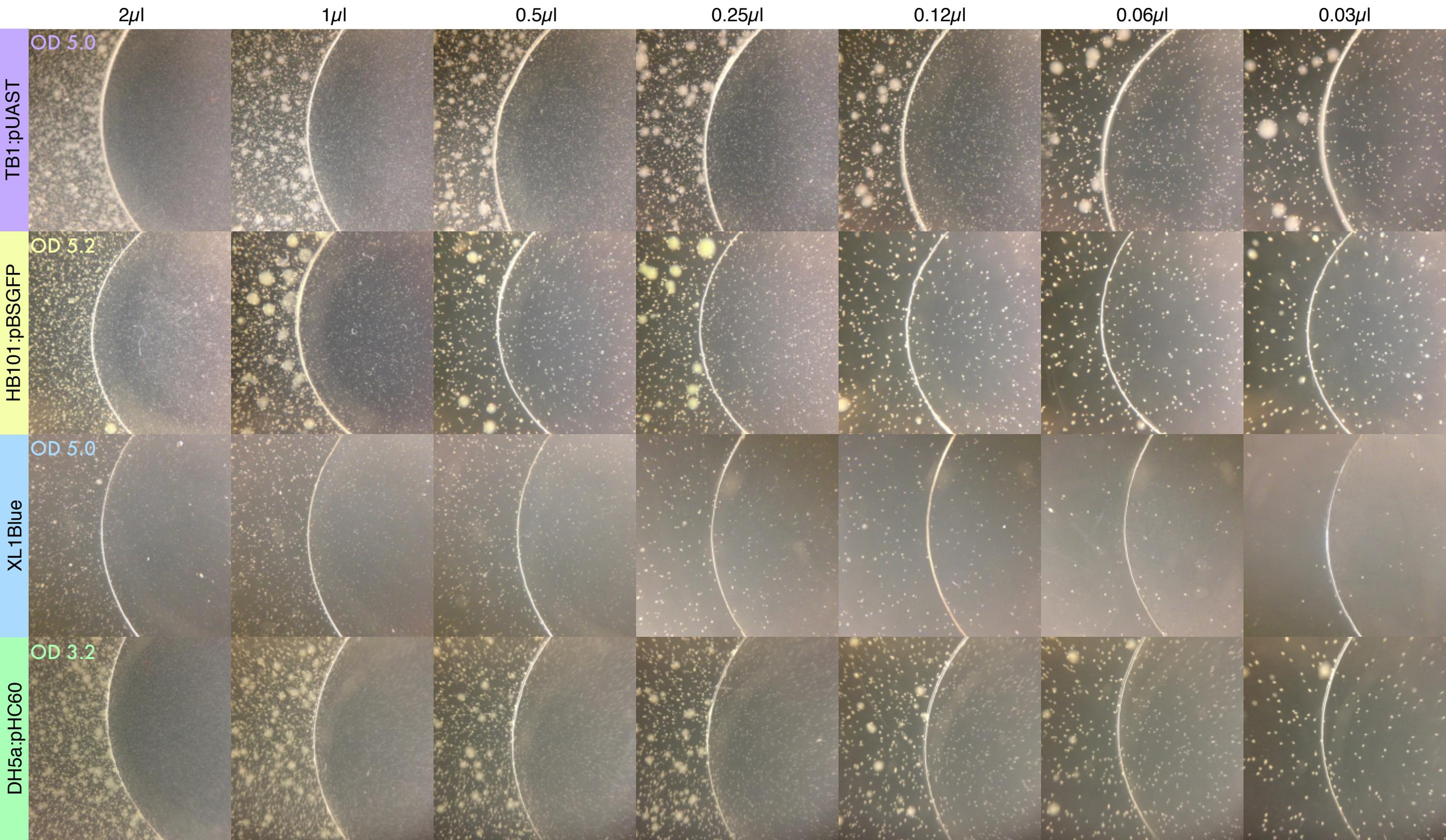

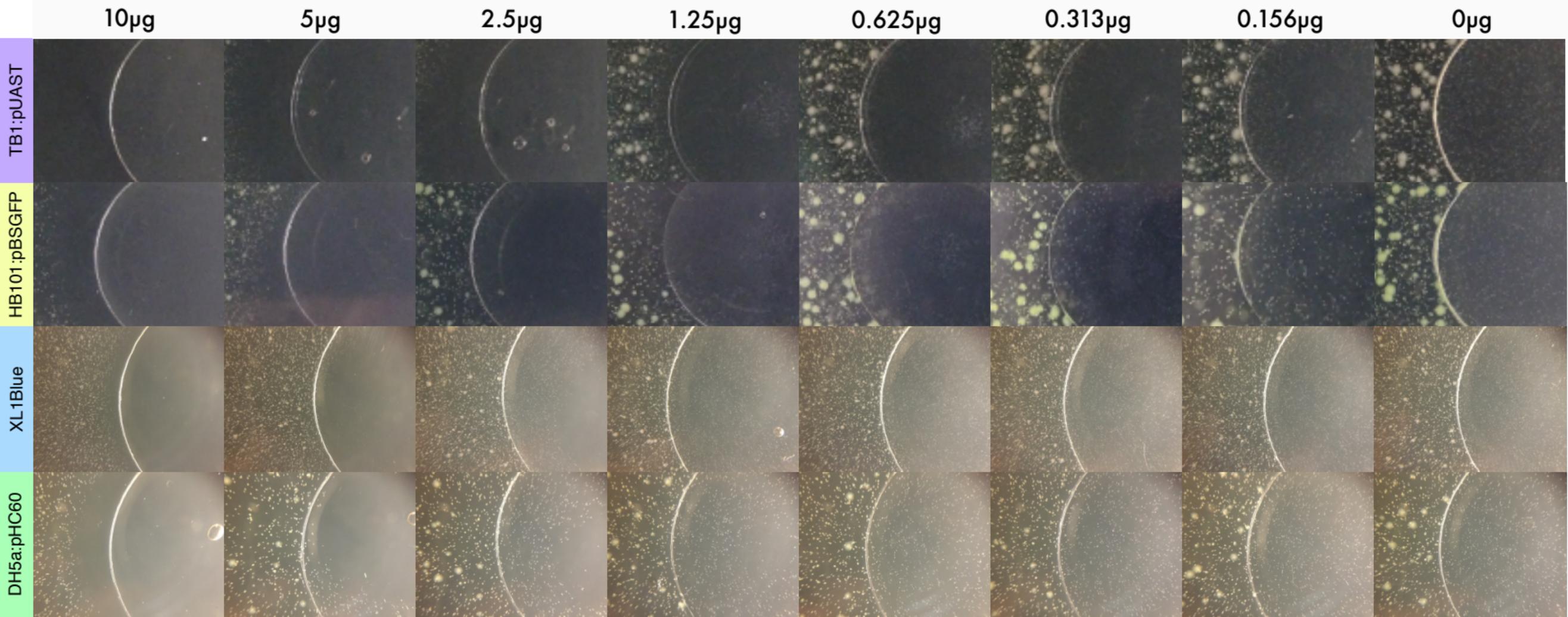

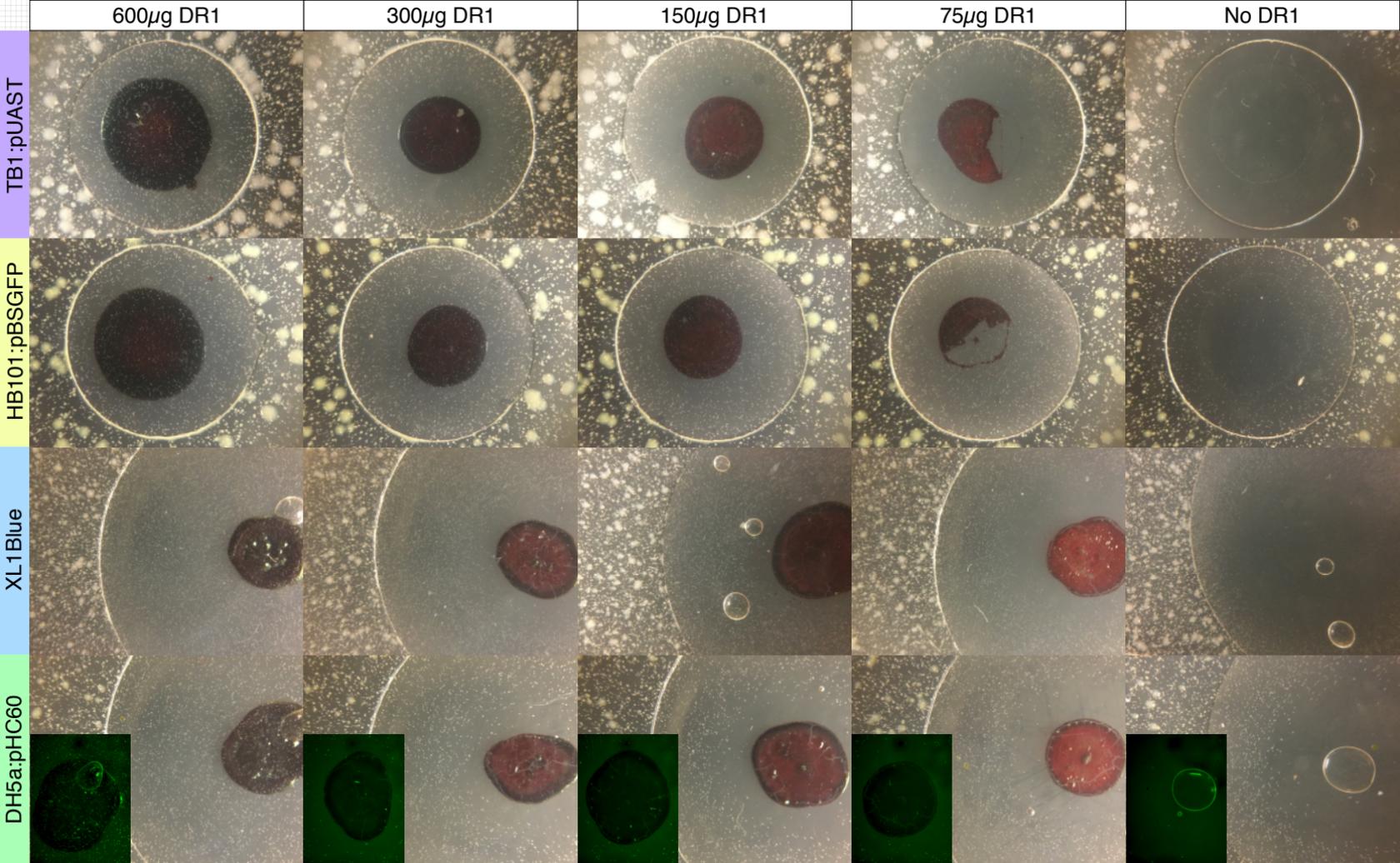

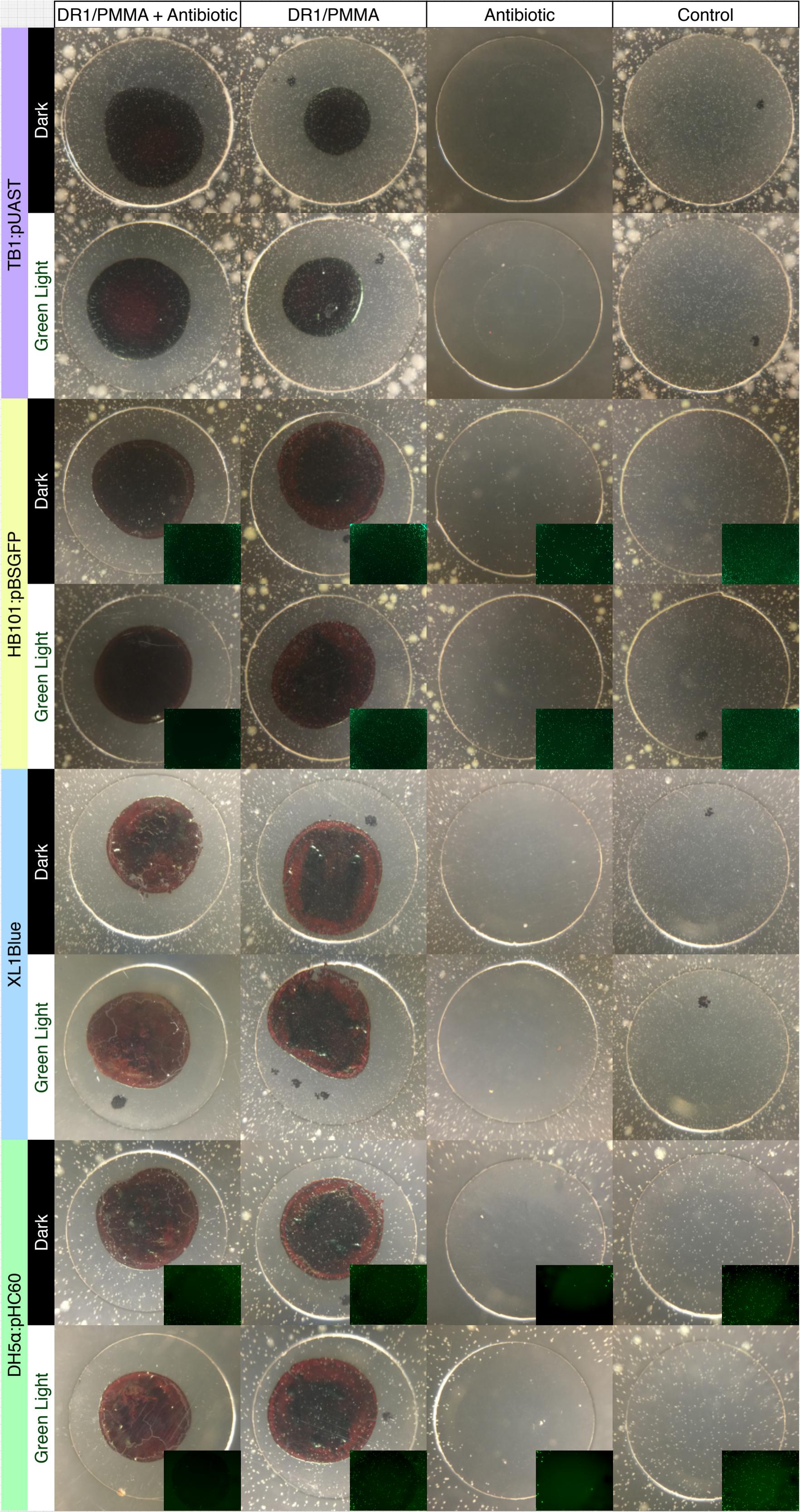

# Supplemental Data

# Capture and light-induced release of antibiotics by an azo dye.

Stephen Atkins[1], Alysa Chueh[2], Taylor Barwell[1], Jean-Michel Nunzi[2] and Laurent Seroude[1]

*Department of Chemistry[2], Department of Biology[1], Queen's University, Kingston, ON, Canada*

## Content

*Capture and light-induced release of antibiotics by an azo dye*

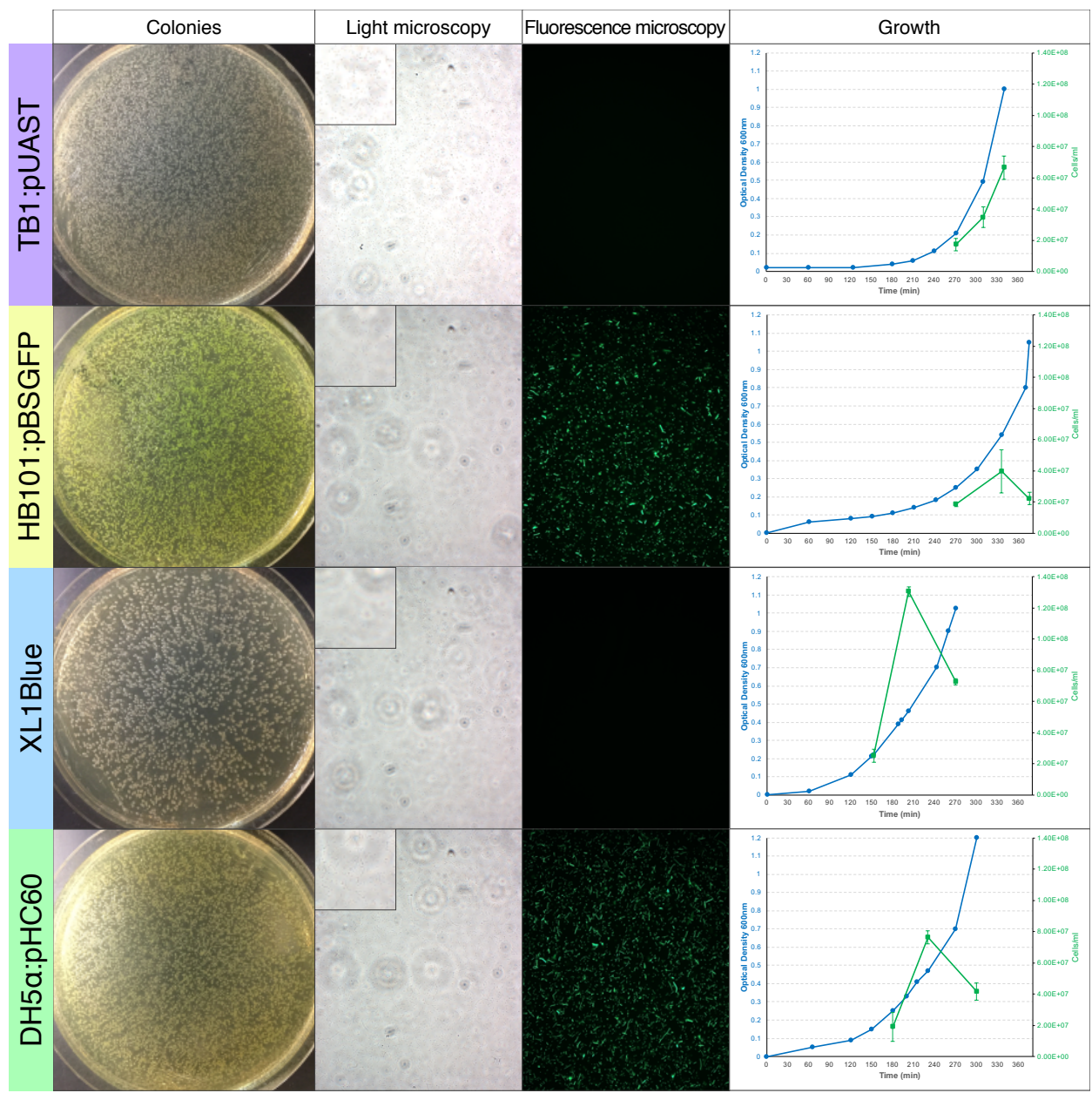

**Figure S1:** *Escherichia coli* **strains used in this study.**

The first column shows the appearance of colonies from each strain growing on 2TY agar. The GFP producing strains HB101:pBSGFP and DH5α:pHC60 exhibit colonies with a different color (green or slightly green). The second column shows the appearance of the cells with a light microscope (400x magnification, inset: digital zoom of subarea). The third column shows the appearance of the same cells with fluorescence microscopy. The fourth column shows the growth kinetics as the change in optical density and cell numbers as a function of time. Strains were cultured in 36ml 2TY supplemented with ampicillin or tetracycline in 250ml flasks under constant shaking in a 37°C water-bath. Cultures were inoculated by 1/100 dilution of an overnight culture. Samples were periodically removed and the optical density at 600nm measured. When OD at 600nm reached approximately 0.2 (early exponential phase), 0.5 (mid-exponential phase) and 1.0 (late exponential phase), three to four samples of the cultures were plated at multiple serial dilutions and the number of colonies obtained after overnight incubation of the plates at 37°C was used to determine the number of viable cells in the culture (error bars: ±SD).

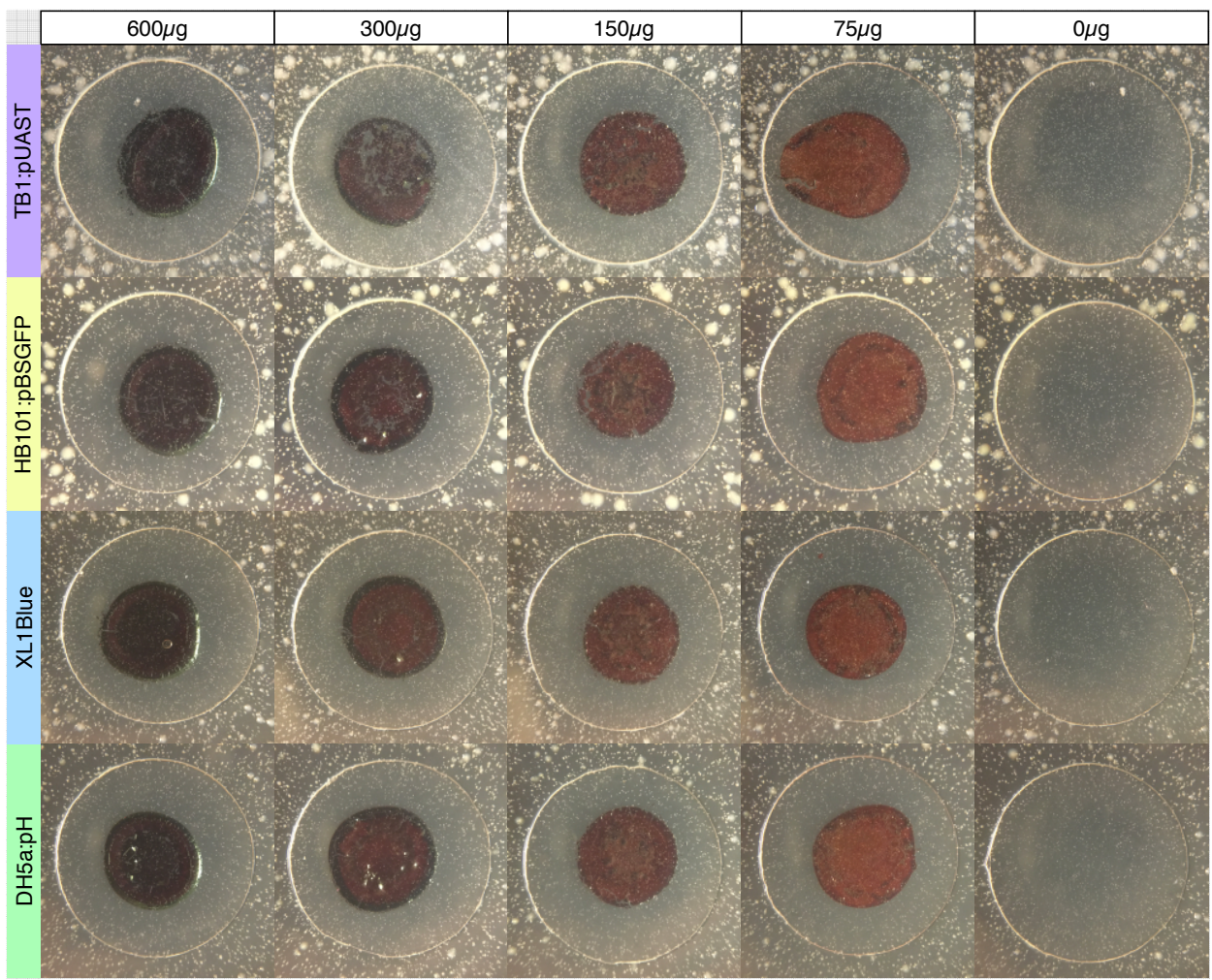

**Figure S2: DR1/PMMA does not affect growth.**

The agar diffusion assay was performed with coverslips with different amounts of DR1/PMMA (indicated at the top of each column). Plates were incubated overnight at 37˚C in the dark. Each strain was tested independently. No strain indicated sensitivity to DR1 at any concentration.

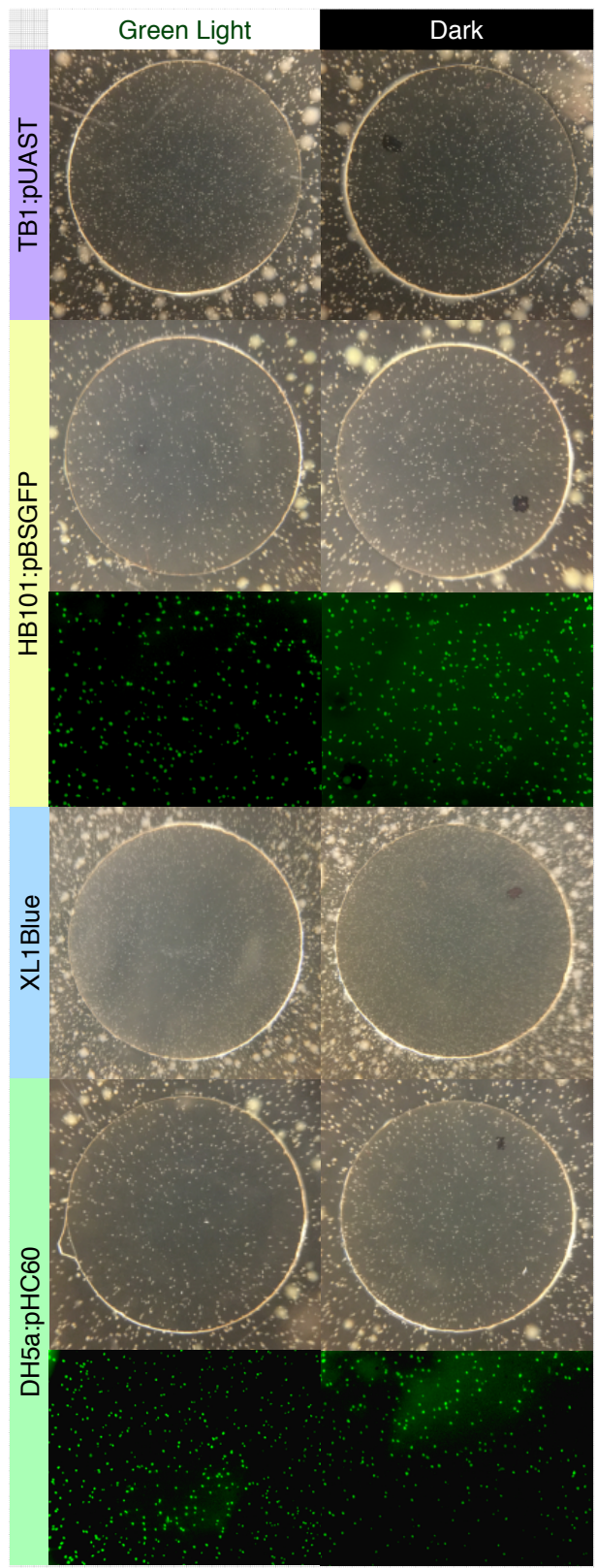

**Figure S3: Green light does not affect growth**

The agar diffusion assay was performed with blank coverslips Effect of green light on colony growth. Plates were incubated overnight in the dark (Dark) or with the green light directly overtop of the coverslip (Green light). The center of the coverslips was also visualized with GFP fluorescence for the GFP producing strains. The green light exposure show none (TB1, DH5α) or negligible effects (XL1Blue, HB101).

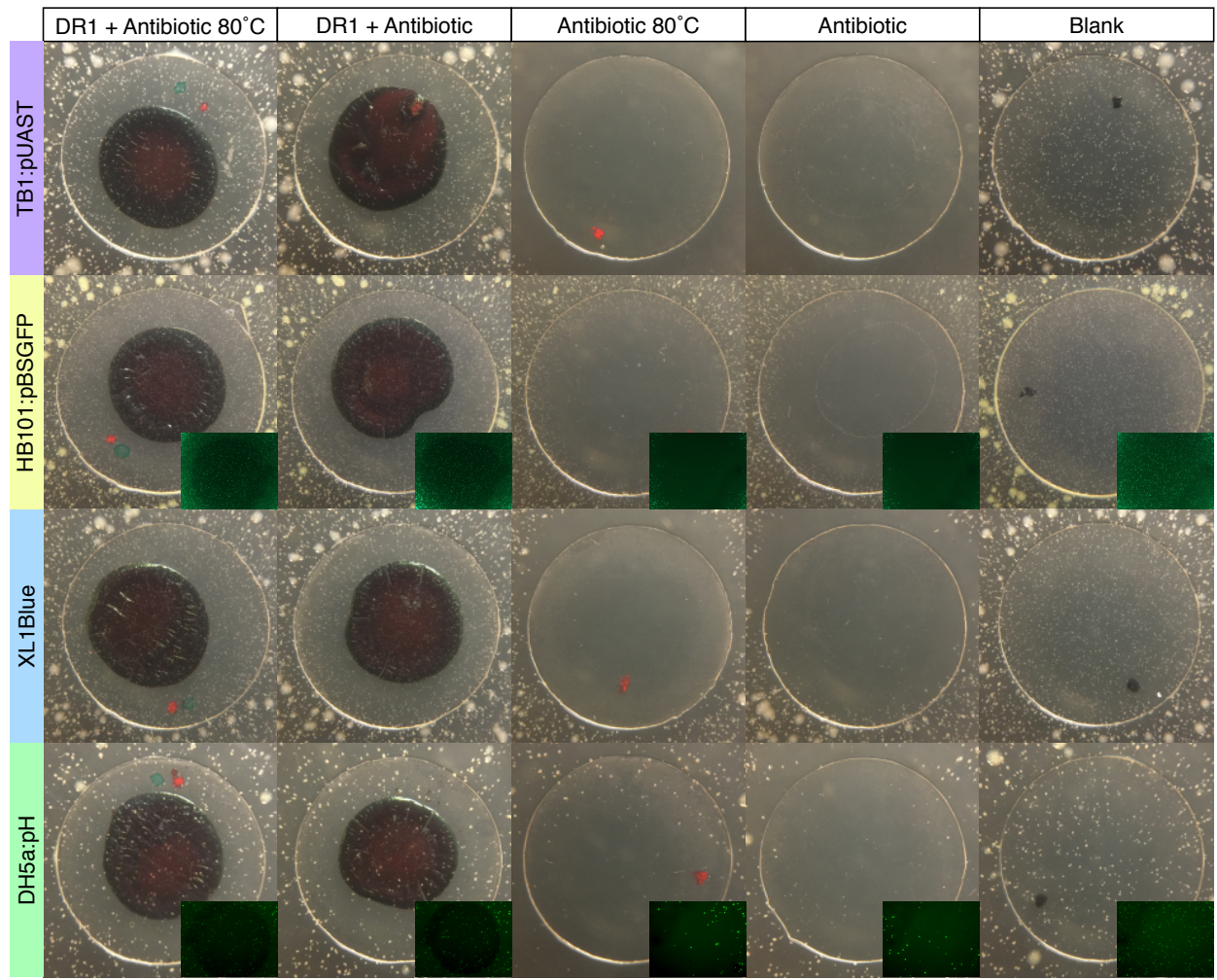

**Figure S4: Antibiotics are not released by heat.**

The agar diffusion assay was performed in the dark with DR1/PMMA + ampicillin (bottom two rows) or + tetracycline (top two rows) coverslips (DR1 + Antibiotic) and positive control coverslips with only ampicillin or tetracycline (Antibiotic). Coverslips were heated at 80˚C for 10 minutes in a vacuum oven and immediately placed atop the agar (80˚C). Antibiotic coverslips were heated and it was observed that ampicillin and tetracycline remained effective after heating. They were plated after cooling for 5 minutes at room temperature. The DR1 plus antibiotic coverslips were applied immediately out of the oven without a cooling period. There is no observable difference between the heated and non-heated coverslips and they are undistinguishable from the negative control coverslip (Blank). The center of the coverslips was also visualized with GFP fluorescence for the GFP producing strains (insets row 2 and 4).

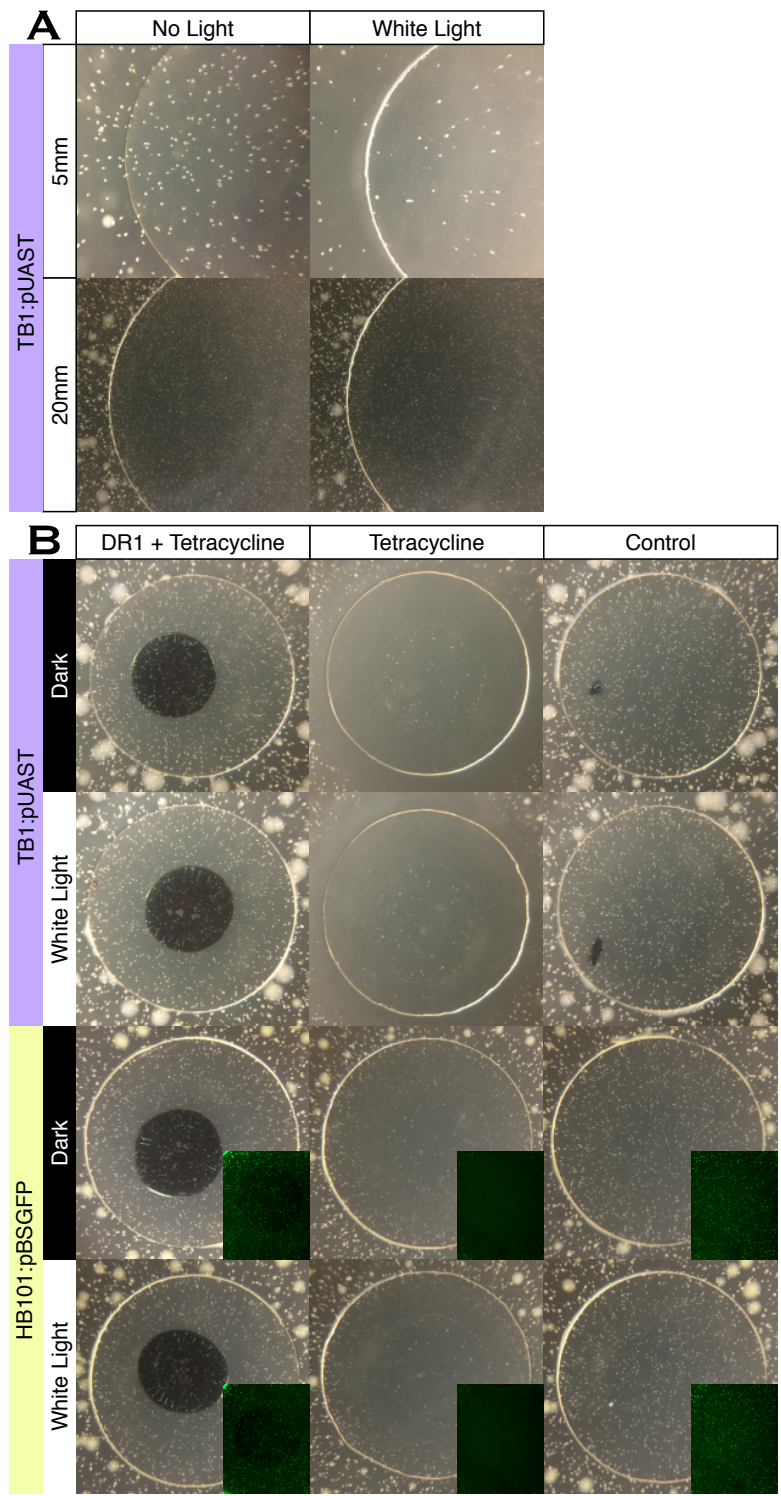

**Figure S5: Tetracyline cannot be released with white light.**
**A.** Unlike green light, white light generates heat that can inhibit bacterial growth. The agar diffusion assay was performed with blank coverslips and was incubated with or without exposure to white light. When the light is placed 5 mm (top) above the coverslip, most of the bacteria are unable to grow. In the next experiment (bottom), the bacterial density was increased and the light placed 20mm above the coverslip. At this distance, the light has no effect. **B.** A release test using coverslips with 600 µg of DR1/PMMA and 0.625 µg of tetracycline was performed with the white light placed 20mm above the coverslips. There is no observable release of tetracycline by exposure to white light.

*Capture and light-induced release of antibiotics by an azo dye*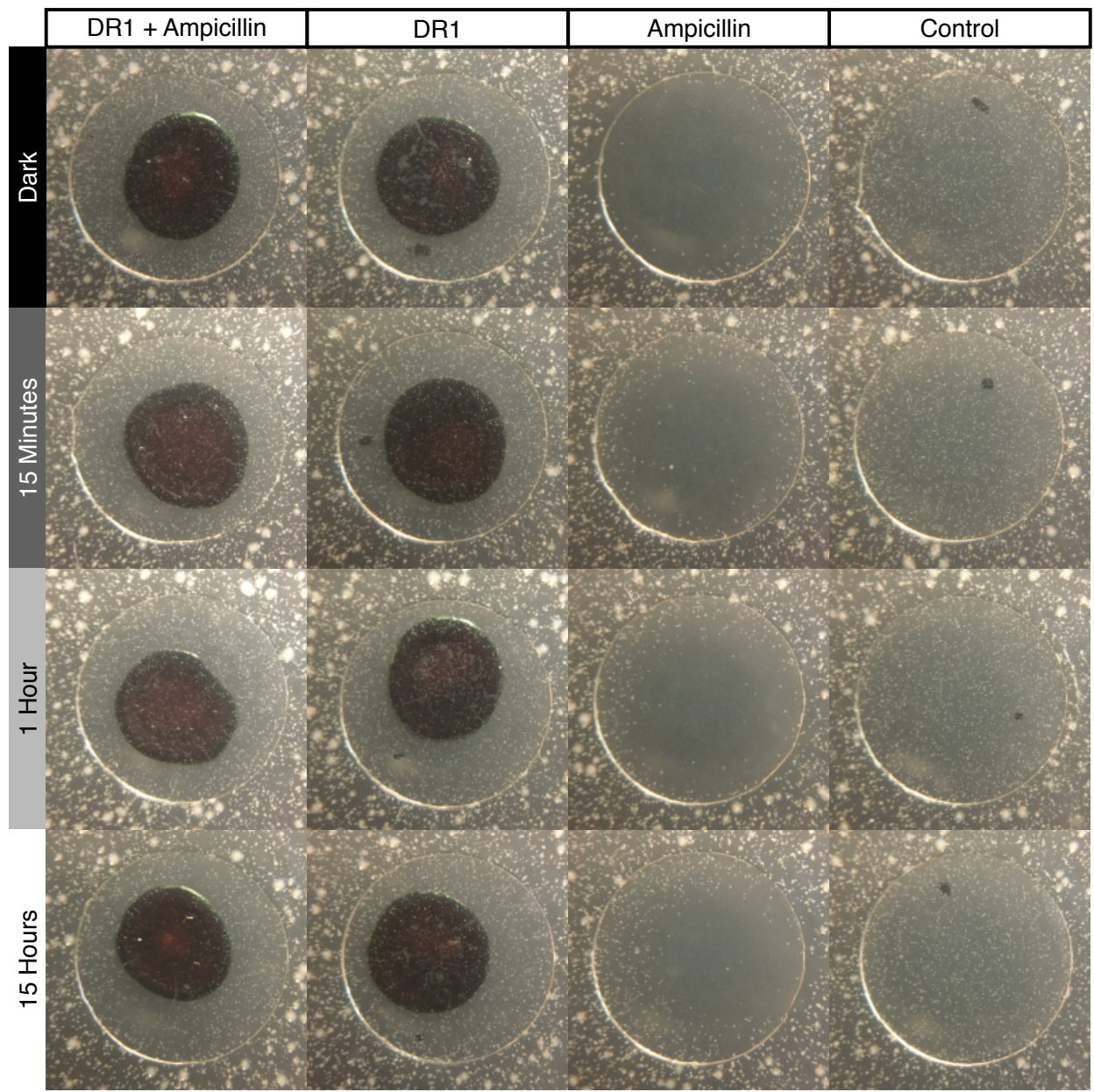

**Figure S6: Timing of release of ampicillin**

Exposure duration required for antibiotic release by DR1.

The agar diffusion assay was performed with the XL1Blue strain. The DR1 + Ampicillin and DR1 coverslips have 600µg DR1/PMMA. The DR1 + Ampicillin and Ampicillin coverslips have 2.5µg ampicillin. The incubation was done without light (Dark) or with the green light atop the coverslips for the indicated amount of time at the beginning of the incubation (15 minutes, 1 hour) and for the whole duration of the incubation (15 hours). The release of ampicillin is not detectable with up to one hour exposure.